\documentstyle[11pt,amssymb]{article}
\newcommand{\be}{\begin{equation}}
\newcommand{\ee}{\end{equation}}
\newcommand{\ba}{\begin{array}}
\newcommand{\ea}{\end{array}}
\newcommand{\bea}{\begin{eqnarray}}
\newcommand{\eea}{\end{eqnarray}}

\renewcommand{\l}{\newline\null}

\newcommand{\rar}{\rightarrow}
\newcommand{\p}{\partial}
\newcommand{\ol}{\overline}

\newcommand{\la}{\langle}
\newcommand{\ra}{\rangle}

\def\hbar{h\!\!\!/}
\begin{document}
\begin{titlepage}
November 1996\hfill PAR-LPTHE 96/45
\vskip 4cm
\centerline{
{\bf \boldmath{$CP$} VIOLATION WITH TWO GENERATIONS}}
\centerline{
{\&}}
\centerline{
{\bf PSEUDOSCALAR MESONS REVISITED.}}
\vskip 5mm
\centerline{B. Machet
     \footnote[1]{Member of `Centre National de la Recherche Scientifique'.}
     \footnote[2]{E-mail: machet@lpthe.jussieu.fr.}
     }
\vskip 5mm
\centerline{{\em Laboratoire de Physique Th\'eorique et Hautes Energies,}
     \footnote[3]{LPTHE tour 16\,/\,1$^{er}\!$ \'etage,
          Universit\'e P. et M. Curie, BP 126, 4 place Jussieu,
          F 75252 PARIS CEDEX 05 (France).}
}
\centerline{\em Universit\'es Pierre et Marie Curie (Paris 6) et Denis
Diderot (Paris 7);} \centerline{\em Unit\'e associ\'ee au CNRS URA 280.}
\vskip 1.5cm
{\bf Abstract:}  After developing in \cite{Machet3} a $SU(2)_L \times U(1)$
gauge theory for $J=0$ mesons, I show in the case of two generations that:\l
-\quad the electroweak mass eigenstates differ, especially in the $K$
and $D$ sector, from what they are assumed to be on the basis of a
classification by the $SU(4)$ group of flavour;\l
-\quad a new mechanism for $CP$ violation springs out, different from
that of the standard model for quarks which requires at least three 
generations,\l
and I construct a $SU(2)_L \times U(1)$ gauge invariant
Lagrangian  which is not invariant by $CP$.
\smallskip

\vfill
\end{titlepage}
\section{Introduction.}

The kaons have a long and rich story of a puzzling and cumbersome system. We
shall not recall here the many steps that went from the $\tau-\theta$ puzzle
\cite{Lee} to the intricacies of $CP$-violation
\cite{ChristensonCroninFitchTurlay}\cite{WinsteinWolfenstein}. 
With the increase of the available energy
came the discovery of the system of the $D$ mesons, similar in many
respects to the latter. Presently, other thresholds have been crossed, but
we shall stay here beyond the bottom threshold
and mainly concentrate on $K$ and $D$ mesons; this corresponds, in the quark
language \cite{GellMann}, to two generations.  
We do not, however, study these particles from the 
quark point of view, but follow  \cite{Machet3} that brings forward 
an electroweak gauge theory for the $J=0$ mesons themselves; those are thus
both the fields and the particles of the model which is, besides, 
compatible with  the electroweak standard model for quarks
\cite{GlashowSalamWeinberg} (see section \ref{section:reps}).

In \cite{Machet3}, I exhibited in particular all their electroweak
representations with a given $CP$ quantum number, in the
form of $N^2/2$ quadruplets ($N/2$ is the number of generations), 
containing a neutral singlet and a triplet of the custodial $SU(2)_V$ 
symmetry; this symmetry, different from what it is generally assumed to be
\cite{SikivieSusskindVoloshinZakharov}, was shown there  to be linked to the 
quantization of
the electric charge (the extension to the leptonic sector was studied in
\cite{Machet4}).  I particularize here the study and show that:\l
-\quad the electroweak mass eigenstates are different from their
usual quark content attributed  from a classification 
by the $SU(4)$ group of flavour, symmetry of strong interactions; in
particular, the alignment of strong and electroweak $K$ and $D$ mesons is
impossible, even at the limit of vanishing mixing (Cabibbo) angle;\l
-\quad unlike what happens in the electroweak standard model for quarks,
\cite{GlashowSalamWeinberg}\cite{KobayashiMaskawa}, one can now construct 
already with two generations a $SU(2)_L \times U(1)$ invariant gauge Lagrangian 
for $J=0$ mesons which is not $CP$ invariant, and the electroweak mass
eigenstates of which are not eigenstates of $CP$.

\section{Electroweak representations of {\boldmath $J=0$} mesons.}
\label{section:reps}

I adopt the point of view that there is no loophole in the
experimental determination of the parity of $J=0$ mesons. We know that this
has been questioned \cite{Lee}, and will see that it might still be, but it 
allows us to work with electroweak mesonic representations the entries of
which have a definite parity, and we put aside in this paper the possibility 
of an admixture of scalars and pseudoscalars. 

We know then from \cite{Machet3} that one can find quadruplet 
representations of the 
electroweak $SU(2)_L \times U(1)$ gauge group which are of two types: the
first contains a scalar singlet and a pseudoscalar triplet of the custodial
$SU(2)_V$:
\be
({\Bbb M}\,^0, \vec {\Bbb M}) = ({\Bbb S}^0, \vec {\Bbb P}),
\label{eq:SP}
\ee
and the second a pseudoscalar singlet and a scalar triplet:
\be
({\Bbb M}\,^0, \vec {\Bbb M}) = ({\Bbb P}\,^0, \vec {\Bbb S}).
\label{eq:PS}
\ee
The $SU(2)_L$ group acts on both by:
\bea
T^i_L. {\Bbb M}^j &=& 
     -{i\over 2}( \epsilon_{ijk}{\Bbb M}^k + \delta_{ij}{\Bbb M}^0),\cr
T^i_L. {\Bbb M}^0 &=&  {i\over 2} {\Bbb M}^j,
\label{eq:action}\eea
and mixes, as expected from its ``left-handed'' nature, scalars and
pseudoscalars.
The action of the $U(1)$ group results from the Gell-Mann-Nishijima
relation as explained in \cite{Machet3}.

All entries are $N\times N$ matrices, where $N$ is
the number of ``flavours'', and the quadruplets can be written

\vbox{
\bea
& &\Phi(\Bbb D)=
({\Bbb M}\,^0, {\Bbb M}^3, {\Bbb M}^+, {\Bbb M}^-)(\Bbb D)\cr =
& & \left[
 {1\over \sqrt{2}}\left(\begin{array}{ccc}
                        {\Bbb D} & \vline & 0\\
                        \hline
                        0 & \vline & {\Bbb K}^\dagger\,{\Bbb D}\,{\Bbb K}
                   \end{array}\right),
{i\over \sqrt{2}} \left(\begin{array}{ccc}
                        {\Bbb D} & \vline & 0\\
                        \hline
                        0 & \vline & -{\Bbb K}^\dagger\,{\Bbb D}\,{\Bbb K}
                   \end{array}\right),
i\left(\begin{array}{ccc}
                        0 & \vline & {\Bbb D}\,{\Bbb K}\\
                        \hline
                        0 & \vline & 0           \end{array}\right),
i\left(\begin{array}{ccc}
                        0 & \vline & 0\\
                        \hline
                        {\Bbb K}^\dagger\,{\Bbb D} & \vline & 0
                    \end{array}\right)
             \right];\cr
& &
\label{eq:reps}
\eea
}

${\Bbb K}$ is a $N/2 \times N/2$ unitary matrix that reduces, for $N=4$ to
the Cabibbo matrix \cite{Cabibbo}; $\Bbb D$ is also a real 
$N/2 \times N/2$ matrix.
That the entries ${\Bbb M}^+$ and ${\Bbb M}^-$ are, up to a sign,
hermitian conjugate ({\em i.e.} charge conjugate), requires that the $\Bbb D$'s 
are restricted to symmetric or antisymmetric matrices.

The group action (\ref{eq:action}) for the quadruplets is a particular case 
of a more general one which involves commuting and anticommuting $N\times N$ 
matrices inside the $U(N)_R \times U(N)_L$ chiral algebra; the generators of 
the $SU(2)_L$ and $U(1)$ subgroups are themselves represented 
by $N\times N$ matrices \cite{Machet3},
\be
{\Bbb T}^3_L = {1\over 2}\left(\begin{array}{ccc}
                        {\Bbb I} & \vline & 0\\
                        \hline
                        0 & \vline & -{\Bbb I}
\end{array}\right),\
{\Bbb T}^+_L =           \left(\begin{array}{ccc}
                        0 & \vline & {\Bbb K}\\
                        \hline
                        0 & \vline & 0           \end{array}\right),\
{\Bbb T}^-_L =           \left(\begin{array}{ccc}
                        0 & \vline & 0\\
                        \hline
                        {\Bbb K}^\dagger & \vline & 0
\end{array}\right),
\label{eq:SU2L}
\ee
acting trivially on $N$-vectors of quarks if they are taken as the
fundamental fields (the action of the gauge group on the mesons can be
deduced from that on the quarks when the former are considered as $\bar q_i
q_j$ or $\bar q_i\gamma_5 q_j$ composite states).
Symmetric $({\Bbb S}^0, \vec {\Bbb P})$ and antisymmetric 
$({\Bbb P}^0, \vec {\Bbb S})$ representations
are $CP$-even, while antisymmetric $({\Bbb S}^0, \vec {\Bbb P})$ and symmetric 
$({\Bbb P}^0, \vec
{\Bbb S})$ representations are $CP$-odd. The $C$, and hence the $CP$  quantum
number of a representation can be swapped by multiplying its entries by $\pm i$.

Restricting to $N=4$, we write below the four types of 
$({\Bbb M}^0, {\Bbb M}^3, {\Bbb M}^+, {\Bbb M}^-)$
quadruplets that appear, corresponding respectively to
\be
{\Bbb D}_1 = {\Bbb I}= \left( \ba{rr} 1 & 0 \cr
                            0 & 1     \ea\right),\quad
{\Bbb D}_2 = \left( \ba{rr} 1 & 0 \cr
                            0 & -1 \ea\right),\quad
{\Bbb D}_3 =  \left( \ba{rr} 0 & 1 \cr
                             1 & 0 \ea\right),\quad
{\Bbb D}_4 =   \left( \ba{rr} 0 & 1 \cr
                             -1 & 0 \ea\right);
\ee
%
%
\vbox{
\bea
& &\Phi({\Bbb D}_1) = \cr
& & \hskip -2cm \left[
{1\over\sqrt{2}} \left(\ba{rrcrr} 1 &   &\vline &    &    \nonumber\\
                                    & 1 &\vline &    &    \nonumber\\
                                    \hline
                                    &   &\vline &  1 &    \nonumber\\
                                    &   &\vline &    &  1 \ea \right),
{i\over\sqrt{2}} \left(\ba{rrcrr} 1 &   &\vline &   &   \nonumber\\
                                    & 1 &\vline &   &   \nonumber\\
                                    \hline
                                    &   &\vline &-1 &   \nonumber\\
                                    &   &\vline &   & -1   \ea \right),
i \left(\ba{rrcrr}   &  &\vline & c_\theta &  s_\theta \nonumber\\
                             &  &\vline &-s_\theta &  c_\theta \nonumber\\
                            \hline
                             &  &\vline &   &     \nonumber\\
                             &  &\vline &   &  \ea \right),
i \left(\ba{rrcrr}   &   &\vline &   &   \nonumber\\
                             &   &\vline &   &   \nonumber\\
                             \hline
                             c_\theta &-s_\theta &\vline &   &
\nonumber\\
                             s_\theta & c_\theta &\vline &   &   \ea \right)
\right]; \nonumber\\
& &
\label{eq:PHI1}
\eea
}
\vbox{
\bea
& &\Phi({\Bbb D}_2) = \cr
& & \hskip -2cm \left[
{1\over\sqrt{2}} \left(\ba{rrcrr}
     1 &   &\vline &    &    \nonumber\\
       & -1 &\vline &    &    \nonumber\\
     \hline
  &  & \vline & c_\theta^2 - s_\theta^2 & 2c_\theta s_\theta  \nonumber\\
  &  & \vline & 2c_\theta s_\theta & s_\theta^2 -c_\theta^2  \ea \right),
{i\over\sqrt{2}} \left(\ba{rrcrr}
             1 &    & \vline &   &   \nonumber\\
               & -1 & \vline &   &   \nonumber\\
             \hline
      &  & \vline & s_\theta^2 - c_\theta^2 & -2c_\theta s_\theta \nonumber\\
      &  & \vline & -2c_\theta s_\theta & c_\theta^2 - s_\theta^2  \ea
\right),
\right .\nonumber\\
& & \hskip 7cm \left .
i \left(\ba{rrcrr}   &  &\vline & c_\theta & s_\theta \nonumber\\
                             &  &\vline & s_\theta & -c_\theta \nonumber\\
                            \hline
                             &  &\vline &   &     \nonumber\\
                             &  &\vline &   &  \ea \right),
i \left(\ba{rrcrr}     &   &\vline &   &   \nonumber\\
                               &   &\vline &   &   \nonumber\\
                              \hline
                              c_\theta & s_\theta &\vline &   & \nonumber\\
                              s_\theta & -c_\theta &\vline &   &   \ea
\right)
\right]; \nonumber\\
& &
\label{eq:PHI2}
\eea
}
\vbox{
\bea
& &\Phi({\Bbb D}_3) = \cr
& & \hskip -2cm \left[
{1\over\sqrt{2}} \left(\ba{rrcrr}
        & 1 &\vline &    &    \nonumber\\
      1 &  &\vline &    &    \nonumber\\
     \hline
     &  & \vline & -2c_\theta s_\theta & c_\theta^2 - s_\theta^2 \nonumber\\
     &  & \vline &  c_\theta^2 - s_\theta^2 & 2c_\theta s_\theta   \ea \right),
{i\over\sqrt{2}} \left(\ba{rrcrr}
         & 1 & \vline &   &   \nonumber\\
       1 &  & \vline &   &   \nonumber\\
       \hline
     &  & \vline & 2c_\theta s_\theta & s_\theta^2 - c_\theta^2 \nonumber\\
     &  & \vline & s_\theta^2 - c_\theta^2 & -2c_\theta s_\theta \ea
\right),
\right .\nonumber\\
& & \hskip 7cm \left .
i \left(\ba{rrcrr}   &  & \vline & -s_\theta & c_\theta \nonumber\\
                             &  & \vline & c_\theta &  s_\theta \nonumber\\
                            \hline
                            &  &\vline &   &     \nonumber\\
                            &  &\vline &   &  \ea \right),
i \left(\ba{rrcrr}     &   &\vline &   &   \nonumber\\
                               &   &\vline &   &   \nonumber\\
                              \hline
                             -s_\theta & c_\theta &\vline &   &
\nonumber\\
                              c_\theta & s_\theta &\vline &   &   \ea
\right)
\right]; \nonumber\\
& &
\label{eq:PHI3}
\eea
}
\vbox{
\bea
& &\Phi({\Bbb D}_4) = \cr
& & \hskip -2cm \left[
{1\over\sqrt{2}} \left(\ba{rrcrr}           & 1 &\vline &    & \nonumber\\
                                         -1 &   &\vline &    & \nonumber\\
                                            \hline
                                            &   &\vline &    & 1 \nonumber\\
                                            &   &\vline & -1 &    \ea
\right),
{i\over\sqrt{2}}\left(\ba{rrcrr}   & 1 &\vline &   &   \nonumber\\
                                         -1 &   &\vline &   &   \nonumber\\
                                           \hline
                                            &   &\vline &   & -1 \nonumber\\
                                            &   &\vline & 1 &    \ea
\right),
 i \left(\ba{rrcrr}   &  &\vline & -s_\theta & c_\theta \nonumber\\
                            &  &\vline & -c_\theta & -s_\theta \nonumber\\
                            \hline
                            &  &\vline &   &     \nonumber\\
                            &  &\vline &   &  \ea \right),
i  \left(\ba{rrcrr}     &   &\vline &   &   \nonumber\\
                              &   &\vline &   &   \nonumber\\
                            \hline
                           s_\theta & c_\theta &\vline &   &   \nonumber\\
                           -c_\theta & s_\theta &\vline &   &   \ea \right)
\right]; \nonumber\\
& &
\label{eq:PHI4}
\eea
}
%
$c_\theta$ and $s_\theta$ stand respectively for the cosine and sine of the
Cabibbo angle $\theta_c$.

We shall also use in the following the notations
\be
({\Bbb S}^0, \vec{\Bbb P})({\Bbb D}_1)= \Phi_1,\quad
({\Bbb S}^0, \vec{\Bbb P})({\Bbb D}_2)= \Phi_2,\quad
({\Bbb S}^0, \vec{\Bbb P})({\Bbb D}_3)= \Phi_3,\quad
({\Bbb S}^0, \vec{\Bbb P})({\Bbb D}_4)= \Phi_4.
\label{eq:notation}\ee
While there is of course no antisymmetric $\Bbb D$ for $N=2$ (one generation),
${\Bbb D}_4$ is such a matrix for $N=4$ and is the only one in this case. 

Though quarks never appear as fundamental fields, the reader can easily
make the link between the mesons, represented above as $4\times 4$ matrices,
and their quark ``content'': it is simply achieved by sandwiching
a given matrix belonging to a representation between fermionic 4-vectors
$(\bar u, \bar c, \bar d, \bar s)$ and $(u,c,d,s)$, and by 
remembering the parity of the corresponding particle.  
With the scaling that has to be introduced \cite{Machet1,Machet2,Machet3}, 
we have, for example
\be
{\Bbb P}^+({\Bbb D}_1) = i{f\over \la H\ra}\left(c_\theta (\pi^+ + D_s^+) +
                                         s_\theta (K^+ -D^+)\right),
\ee
where, according to the classification by flavour $SU(4)$, we have translated,
for pseudoscalars, $\bar u  d$ into  $\pi^+$, $\bar u s$ into 
$K^+$, $\bar c d$ into $D^+$ and $\bar c s$ into $D_s^+$ etc \ldots;  
$f$ is the leptonic decay constant of the mesons (considered to be the same 
for all of them) and $H$ is the Higgs boson.

We always refer to $\pi, K, D, D_s$ as the eigenstates of strong interactions.

\section{The mass and \boldmath{$CP$} eigenstates for
\boldmath{$\theta_{c}=0$}.}

\subsection{Quadratic invariants.}

For the sake of simplicity, we shall deal here with the case of vanishing 
Cabibbo angle. It teaches us the main features of the unavoidable
misalignment between strong and electroweak eigenstates.

To every representation is associated a quadratic expression invariant
by the electroweak gauge group
\be
{\cal I} = ({\Bbb M}^0, \vec {\Bbb M})\otimes ({\Bbb M}^0, \vec {\Bbb M})=
 {\Bbb {\Bbb M}}\,^0 \otimes {\Bbb {\Bbb M}}\,^0 +
                 \vec {\Bbb M} \otimes \vec {\Bbb M};
\label{eq:invar}
\ee
the ``$\otimes$'' product is a tensor product, not
the usual multiplication of matrices and means the product 
of fields as functions of space-time; $\vec {\Bbb M} \otimes \vec {\Bbb M}$ 
stands for $\sum_{i=1,2,3} {\Bbb M}\,^i \otimes  {\Bbb M}\,^i$.

The representations $\Phi$ are such that the algebraic sum (to be specified
below) of the corresponding
invariants is diagonal both in the electroweak basis and in the basis of
strong eigenstates; in particular, the (quadratic) part of the kinetic terms 
that involves ordinary derivatives
\bea
{1\over 2} \sum_{i= 1,2,3}& &\left(
-\p_\mu ({\Bbb S}^0, \vec {\Bbb P})({\Bbb D}_i)
                     \otimes \p^\mu ({\Bbb S}^0, \vec {\Bbb P})({\Bbb D}_i)+
\p_\mu ({\Bbb S}^0, \vec {\Bbb P})({\Bbb D}_4)
              \otimes \p^\mu ({\Bbb S}^0, \vec {\Bbb P})({\Bbb D}_4)\right.\cr 
& &\left.
+\p_\mu ({\Bbb P}^0, \vec {\Bbb S})({\Bbb D}_i)
                     \otimes \p^\mu ({\Bbb P}^0, \vec {\Bbb S})({\Bbb D}_i)
-\p_\mu ({\Bbb P}^0, \vec {\Bbb S})({\Bbb D}_4)
                 \otimes \p^\mu ({\Bbb P}^0, \vec {\Bbb S})({\Bbb D}_4)\right)
\label{eq:LK}\eea
is also diagonal in the strong eigenstates, with the same normalization factor 
$1$ for all of them; the relative signs that must be introduced 
for this purpose are due to the following:\l
- all pseudoscalars we define without an ``$i$'' 
(like $\pi^+ = \bar u d_{P-odd}$),
such that the $({\Bbb P}^0, \vec{\Bbb S})$ quadruplets have to be multiplied 
by $\pm i$;\l
- the skew-symmetry of ${\Bbb D}_4$ making the corresponding quadruplets 
have an opposite behaviour by charge conjugation as compared to the other six,
introduces an extra minus sign in the corresponding quadratic invariants.

Another invariant is the
``scalar product'' of two representations transforming alike by the gauge
group; for example such is
\be
{\cal I}_{12} = \Phi_1 \otimes \Phi_2
 ={\Bbb S}^0({\Bbb D}_1) \otimes {\Bbb S}^0({\Bbb D}_2) + 
                  \vec {\Bbb P}({\Bbb D}_1) \otimes \vec {\Bbb P}({\Bbb D}_2).
\label{eq:I12}\ee
Because of the remark starting section \ref{section:reps}, we {\em a priori} 
exclude connecting mesons of different parities inside an invariant like
$({\Bbb S}^0, \vec {\Bbb P})({\Bbb D}_1)\otimes  ({\Bbb P}^0, \vec {\Bbb
S})({\Bbb D}_2)= {\Bbb S}^0({\Bbb D}_1) \otimes {\Bbb P}^0({\Bbb D}_2) + 
                 \vec {\Bbb P}({\Bbb D}_1) \otimes \vec {\Bbb S}({\Bbb D}_2)$.

Note that the $SU(2)_L \times U(1)$ invariants do {\em not} involve tensorial 
products of hermitian conjugate (charge conjugate) fields, but of the fields 
themselves; for example ${\cal I}_{12} =\Phi_1 \otimes \Phi_2$ is given by 
(\ref{eq:I12}) and  {\em not} by
${\Bbb S}^0 \otimes {\Bbb S}^{0\dagger} 
                + \vec{\Bbb P} \otimes \vec{\Bbb P}^\dagger$.
This underlies the results below.

\subsection{A first possible attitude.}\label{subsection:first}

One can choose the electroweak mass eigenstates to match the quadruplets 
displayed above; one then does not introduce crossed mass terms, and we have 
{\em a priori} eight independent mass scales. The mass eigenstates are also
$CP$ eigenstates, but it is obvious that electroweak and strong eigenstates 
differ.

Presumably, this choice is only reasonable in the case of three generations, 
because
of the content of $\Phi_1$ \cite{Machet3}: $\Phi_1^0$ is the Higgs boson and
the three $\vec\Phi$ form the triplet of Goldstone bosons of the broken
electroweak gauge group; once ``eaten'' by the three gauge fields that
get massive by so doing, they become their longitudinal degrees of freedom,
the mass of which are consequently expected to match those of the gauge
bosons; only in the case of three generations can we expect three (at least)
pseudoscalar mesons to be as heavy as the $W^\pm, Z$. We know now that some
mesons interpreted to contain the ``top'' quark weight as much as $175 GeV$
\cite{topmass},
but the possibility is wide open that three among the eleven pseudoscalar
mesons including the top quark are identical with the longitudinal $W^\pm,
Z$: indeed, in the present picture, the mass of the asymptotic states
(mesons) is disconnected from that of their constituents fields, and
different mass scales can  be attributed, in a $SU(2)_L \times
U(1)$ invariant way, to different representations. independently of their
``quark content''; since the eleven above mentioned ``topped'' pseudoscalar 
mesons  will fit into several quadruplets, there is no reason why they
should correspond to a single mass scale.

\subsection{The \boldmath{$\pi-D_s$} mass splitting.}

The second attitude is to attempt to align strong and weak eigenstates, at
least some of them,  and we first focus on the two representations 
$\Phi_1$ and $\Phi_2$ defined in (\ref{eq:notation}).

Consider the mass term
\be
{\cal L}_m  =  {1\over 2} ( m_1^2 \Phi_1 \otimes \Phi_1
                                    + m_2^2  \Phi_2 \otimes \Phi_2
                                    -2m_{12}^2  \Phi_1 \otimes \Phi_2).
\ee
It is diagonalized, in the charged sector for example, by the states
\bea
A^+ &=&{1\over 2}\left(
\sqrt{m_1\over m_2}(\pi^+ + D_s^+) + \sqrt{m_2\over m_1}(\pi^+ - D_s^+)
\right),\cr
B^+ &=&{1\over 2}\left(
\sqrt{m_1\over m_2}(\pi^+ + D_s^+) - \sqrt{m_2\over m_1}(\pi^+ - D_s^+)
\right),
\eea
and their charge conjugate. The same happens in the neutral sector.

For $m_1 = m_2 =m$, the kinetic terms are also diagonal in 
$\vec A$ and $\vec B$,
in which case $\vec A$ and $\vec \pi$ are aligned, so are $\vec B$ and 
$\vec D_s$, with masses squared $m^2 \pm m_{12}^2$.

\subsection{The \boldmath{$K-D$} system.}\label{subsection:KD}

We shall now see that,  because the matrices ${\Bbb D}_3$ and ${\Bbb D}_4$ 
have opposite symmetry properties, it is impossible to align strong and 
electroweak eigenstates in the $K-D$ system (the only exception is the
trivial one, corresponding to degenerate $K$ and $D$ mesons).

Consider the two electroweak representations $U$ and $V$ obtained by
combining $\Phi_3$ and $\Phi_4$ defined in (\ref{eq:notation})

\vbox{
\be
\left\{
\ba{ccc}
\Phi_3 &=& \alpha U + \beta V, \\
\Phi_4 &=& \delta U + \zeta V.
\ea
\right.
\label{eq:PhiPhi}\ee
}

That the ($CP$ invariant) kinetic term 
\be
{\cal L}_{kin} =-{1\over 2}(\p_\mu \Phi_3 \otimes\p^\mu \Phi_3
                                 - \p_\mu \Phi_4 \otimes\p^\mu \Phi_4)
\label{eq:Lkin}
\ee
stays diagonal in $U$ and $V$ requires
\be
\alpha\beta -\delta\zeta =0.
\label{eq:diag}\ee
Let us introduce the ($CP$ invariant) mass terms
\be
{\cal L}_m = {1\over 2} (m_3^2 \Phi_3 \otimes \Phi_3
        -m_4^2 \Phi_4 \otimes \Phi_4
       - 2im_{34}^2 \Phi_3 \otimes \Phi_4).
\label{eq:Lm}\ee
We leave aside the case $m_3^2 = m_4^2,\  m_{34}^2 =0$ which
corresponds to degenerate $K$ and $D$.

If (\ref{eq:Lm}) can be diagonalized together with the kinetic terms, 
there should
exist two mass scales $\mu_U^2$ and $\mu_V^2$ such that  the quadratic
Lagrangian invariant by $SU(2)_L \times U(1)$ is diagonal in $U$ and $V$:
it then reads, using the condition (\ref{eq:diag})
\be
{\cal L} ={1\over 2}(\delta^2 -\alpha^2)
    (\p_\mu U\otimes \p^\mu U -{\beta^2\over \delta^2} \p_\mu V\otimes
\p^\mu V)
         -{1\over 2}(\mu_U^2 U\otimes U -\mu_V^2 V\otimes V).
\label{eq:LUV}\ee
The masses of $U$ and $V$ are
\be
m_U^2 = {1\over \delta^2 -\alpha^2}\mu_U^2,\quad
m_V^2 = {\delta^2 \over \beta^2(\delta^2 -\alpha^2)} \mu_V^2.
\ee
We shall take hereafter, without loss of generality
\be
\delta^2 -\alpha^2 =1
\label{eq:cond}\ee
and look for real $\delta^2$ and $\alpha^2$. The condition of reality is not a
restriction for what we look at since  $U$ and $V$ can only be 
aligned with strong eigenstates for $\alpha$ and $\delta$ both real or
both imaginary (see (\ref{eq:UV})); this is impossible as we show below.

Still making use of the condition (\ref{eq:diag}), and of the relation
(\ref{eq:cond}), eqs.~(\ref{eq:PhiPhi}) 
invert to
\be
\left\{
\ba{ccc}
U &=&   \delta \Phi_4 -\alpha \Phi_3,\cr
V &=& {\delta\over\beta}(\delta\Phi_3 -\alpha\Phi_4).
\ea
\right.
\label{eq:UV}\ee
Replacing in (\ref{eq:LUV}), one gets, by matching it with (\ref{eq:Lm}), the
system

\vbox{
\be
\left\{
\ba{ccccc}
m_3^2 &=& -\alpha^2 \mu_U^2 +\delta^2{\delta^2\over\beta^2}\mu_V^2
      &=& \mu_U^2  - \delta^2(\mu_U^2 - {\delta^2\over\beta^2}\mu_V^2),\cr
m_4^2 &=& \phantom{-}\delta^2 \mu_U^2 -\alpha^2 {\delta^2\over\beta^2}\mu_V^2
      &=& \mu_U^2 + \alpha^2(\mu_U^2 - {\delta^2\over\beta^2}\mu_V^2),\cr
m_{34}^2 &=& i\alpha\delta(\mu_U^2 - {\delta^2\over\beta^2}\mu_V^2), & &
\ea
\right. .
\label{eq:mmm}\ee
One can extract from the first two equations of (\ref{eq:mmm}):
\be
m_U^2 = \frac{m_4^2 +\alpha^2(m_3^2 + m_4^2)}{1 +2\alpha^2},\quad
m_V^2 = \frac{m_3^2 +\alpha^2(m_3^2 + m_4^2)}{1 +2\alpha^2},
\label{eq:mUmV}\ee
}

and, from the third equation of (\ref{eq:mmm}) together with (\ref{eq:cond}),
one finds that  $\alpha^2$ must satisfy
\be
\alpha^4 +\alpha^2 +{\xi^2 \over 4(1+\xi^2)} =0,\quad
\xi = {2m_{34}^2 \over m_4^2 - m_3^2}.
\label{eq:alpha1}\ee
It only has real solutions in $\alpha ^2$ for $\alpha^2 < 0$; we then go to
$\rho = -i\alpha$, and the solutions of (\ref{eq:alpha1}) are
\be
\rho^2 = \frac{1 \pm\sqrt{1\over 1+\xi^2}}{2}.
\label{eq:rho}
\ee
$\rho^2$ is always smaller than $1$, such that $\delta^2 = 1-\rho^2$ is
always positive and $\delta$ real. $U$ has $CP = -1$, and $V$ has $CP = +1$
if $\beta$ is chosen to be real.

Due in particular to $\alpha = i\rho$ in eqs.~(\ref{eq:UV}), both $U$ and $V$
are different from ``strong'' eigenstates. The case $\rho =0$
corresponds to a vanishing crossed mass term $m_{34}^2 =0$ and to the
alignment of $U$ with $\Phi_3$, and of $V$ with $\Phi_4$ (see subsection
\ref{subsection:first}).

The apparently singular case $\alpha^2 =-1/2$ in (\ref{eq:mUmV}) is better
treated directly from eq.~(\ref{eq:Lm}) since it also corresponds to
$m_3^2 = m_4^2 =m^2$: the eigenvectors are $\Phi_3 \pm i\Phi_4$ and the
corresponding masses $m^2 \pm m_{34}^2$.

To fix the ideas, let us take a very simple example:
$\xi^2 = 3, \rho = 1/2, \delta =\sqrt{3}/2$; one has then
\be
m_U^2 = {3 m_4^2 - m_3^2 \over 2}, \quad
m_V^2 = {3 m_3^2 - m_4^2 \over 2};
\ee
$U^\pm$ and $V^\pm$ write

\vbox{
\bea
U^+ &=& \hphantom{-}{1\over 2}\left((1+i\sqrt{3})K^+  
                                   + (1-i\sqrt{3})D^+\right),\cr
U^- &=& {1\over 2}\left((1-i\sqrt{3})K^- +(1+i\sqrt{3})D^-\right) =\ol{U^+}
=-CP\quad U^+,\cr
V^+ &=& {\sqrt{3}\over 4\beta}
              \left((1+i\sqrt{3})K^+ -(1-i\sqrt{3})D^+\right),\cr
V^- &=& -{\sqrt{3}\over 4\beta}
              \left((1-i\sqrt{3})K^- -(1+i\sqrt{3})D^-\right)=-\ol{V^+}
= +CP\quad V^+.
\eea
}

Since $iK^+$  and  $K^+$ have opposite $CP$ transformation, and likewise 
$iD^+$ and $D^+$, the charged electroweak mass eigenstates are expected to 
decay into two as well as three pions. This provides a natural explanation 
for the $\tau-\theta$ puzzle in the charged sector \cite{Lee}.

In the neutral sector, one gets:

\vbox{
\bea
U^3 &=& {1\over 2\sqrt{2}}\left((1+i\sqrt{3})(\ol{D^0}-K^0)
      +(1-i\sqrt{3})({D^0}-\ol{K^0})\right) = +\ol{U^3}= -CP\quad U^3,\cr
V^3 &=& {\sqrt{3}\over 4\sqrt{2}\beta}\left((1+i\sqrt{3})(\ol{D^0}-K^0)
      -(1-i\sqrt{3})({D^0}-\ol{K^0})\right) =-\ol{V^3}= +CP\quad V^3.\cr
& &
\eea
}
$U^3$ and $\pm iU^3$ have opposite $C$, thus opposite $CP$, and are degenerate 
in mass; so are $V^3$ and $\pm iV^3$; 
in fact, because of the $C$ quantum number, we do  not deal, for $N=4$, with
sixteen pseudoscalar mesons, but with twice as many; they are pairwise
degenerate when the mass eigenstates are also $CP$ eigenstates. The same
occurs in the scalar sector.
 
That the lightest pair, for example $U^3, \pm iU^3$ could be identified as
the short-lived and long-lived neutral electroweak kaons is left as an open
possibility.
\footnote{In \cite{Machet3}, I proposed to look for $K_L$ and $K_S$
respectively as the neutral ${\Bbb P}^3$ of a $({\Bbb S}^0, \vec{\Bbb P})$
representation and the ${\Bbb P}^0$ of a $({\Bbb P}^0, \vec{\Bbb S})$.
That one is then at a loss to explain the near mass degeneracy between the
two motivates the alternate proposition made above.}

\section{\boldmath{$CP$} violation with two generations 
          (case \boldmath{$\theta_c =0$}).}

We have all the necessary ingredients to construct an $SU(2)_L \times U(1)$
invariant Lagrangian which is not $CP$ invariant: hence, the corresponding mass
eigenstates do not have a definite $CP$; it only requires the
existence of at least one antisymmetric $\Bbb D$ matrix, that is two
generations.
The principle is to find eigenstates which diagonalize the entire
quadratic Lagrangian, but which are linear combinations of $\Phi_i, i=1,2,3$ 
and $\Phi_4$ with {\em real} coefficients: the two types of quadruplets
having different $CP$ properties, the eigenstates will not have a
definite transformation by this operation.

We shall work again, for example, in the mesonic subspace spanned by the two 
representations $\Phi_3$ and $\Phi_4$, but it must be clear that the
phenomenon is
more general and can occur with any set of quadruplets corresponding
to two $\Bbb D$ matrices ${\Bbb D}_i, i=1,2,3$ and ${\Bbb D}_4$.

With the same notations as in the previous section, we introduce the
$SU(2)_L\times U(1)$ invariant, but now $CP$ non-invariant (without ``$i$''
in the crossed mass term) quadratic mass Lagrangian
\be
{\cal L}_m = {1\over 2}( m_3^2 \Phi_3\otimes \Phi_3
                         -m_4^2 \Phi_4\otimes \Phi_4
                        + 2m_{34}^2 \Phi_3\otimes \Phi_4).
\ee
The $CP$ invariant kinetic terms (\ref{eq:Lkin}) we keep unaltered, though
expressed in terms of $U$ and $V$; that they stay diagonal as before
requires again  that the condition (\ref{eq:diag}) be satisfied.

The argument goes exactly along the same lines as in the previous
section, except that eq.~(\ref{eq:alpha1}) is now replaced by
\be
\alpha^4 + \alpha^2 -{\xi^2 \over 4(1-\xi^2)} =0,\quad
\xi = {2m_{34}^2 \over m_4^2 - m_3^2},
\ee
and has for solution
\be
\alpha^2 = {-1 \pm \sqrt{1\over 1-\xi^2}\over 2}.
\ee
The existence of a positive real solution for $\alpha^2$ requires
$\xi^2 <1$, that is $2m_{34}^2 \leq \vert m_4^2 -m_3^2 \vert$.

To fix the ideas as in the previous section, let us take the simple
example $\xi^2 = 3/4, \alpha = \sqrt{1/2}, \delta =\sqrt{3/2}$. One has
\bea
U^+ &=& -{i\over\sqrt{2}}\left((1 -\sqrt{3})K^+ +(1 +\sqrt{3})D^+\right),\cr
U^- &=& -{i\over\sqrt{2}}\left((1+\sqrt{3})K^- +(1-\sqrt{3})D^-\right)
                 \not = \pm CP\quad U^+,\cr
V^+ &=& -{i\sqrt{3}\over 2\beta}
                \left((1-\sqrt{3})K^+ - (1+\sqrt{3})D^+\right),\cr
V^- &=& \hphantom{-}{i\sqrt{3}\over 2\beta}
                \left((1+\sqrt{3})K^- -(1-\sqrt{3})D^-\right)
                  \not = \pm CP\quad V^+,
\eea
and, in the neutral sector,

\vbox{
\bea
U^3 &=& -{i\over 2}\left((1-\sqrt{3})(\ol{D^0}-K^0) 
                        + (1+\sqrt{3})(D^0 -\ol{K^0})\right),\cr
V^3 &=& -{i\sqrt{3}\over 2\sqrt{2}\beta}\left((1-\sqrt{3})(\ol{D^0}-K^0) 
                        - (1+\sqrt{3})(D^0  -\ol{K^0})\right).
\eea
}

$\ol{U^3} \not = \pm U^3,\ \ol{V^3} \not = \pm V^3$: the mass eigenstates
$U^3$ and $V^3$ are consequently not $CP$ eigenstates.

The masses of $\vec U$ and $\vec V$ are
\be
m_U^2 = {3m_4^2 + m_3^2\over 4},\quad m_V^2 = {3m_3^2 + m_4^2\over 4}.
\ee

Treating in perturbation  a Lagrangian like (\ref{eq:LUV}) requires being able 
to use Green functions of the form $\la T \varphi(x) \bar\varphi(y) \ra$. 
When $U$ and $V$ were $C$ (and $CP$) eigenstates as in the previous section, 
all their entries were related to their charge conjugate by a simple sign, but
this is no longer the case . For this reason,
one must now switch to the fields $A=(U+\bar U)/2, B=(U -\bar U)/2,
C= (V+\bar V)/2, D= (V-\bar V)/2$, the entries of which have definite 
properties by charge conjugation; this transforms (\ref{eq:LUV}) into

\vbox{
\bea
{\cal L} =& &{1\over 2}(\delta^2 -\alpha^2)(
  (\p_\mu A \otimes\p^\mu\bar A -\p_\mu B \otimes\p^\mu\bar B
           + 2 \p_\mu A \otimes \p^\mu B) \cr
& &   - {\beta^2\over \delta^2} (\p_\mu C \otimes\p^\mu\bar C 
                                        -\p_\mu D \otimes\p^\mu\bar D
                                     +2 \p_\mu C\otimes \p^\mu D))\cr
  &-&{1\over 2}\left(\mu_U^2 (A\otimes \bar A - B\otimes \bar B +2 A\otimes B)
             -\mu_V^2 (C\otimes \bar C - D\otimes \bar D +2 C\otimes D)\right);
\eea
}

though the ``masses'' stay unchanged, unavoidable non-diagonal quadratic terms,
including derivative ones, appear.

The whole kinetic Lagrangian, obtained from (\ref{eq:LK}) by replacing normal
derivatives with covariant derivatives  with respect to $SU(2)_L \times U(1)$,
is $CP$ invariant.
But while, by the construction given above, the part that only 
includes normal derivatives can be diagonalized either in $\vec\Phi_3$ and 
$\vec\Phi_4$ or in $\vec U$ and $\vec V$, there is no reason why the same 
should happen 
for the remaining terms which couple to the electroweak gauge bosons.
As a consequence, loop corrections with gauge fields
are expected to induce transitions between the $U$ and $V$ mass eigenstates,
and the independent symmetries $U \rar \pm iU, V \rar \pm iV$ mentioned at the
end of subsection \ref{subsection:KD} are broken.

\section{The case of non-vanishing Cabibbo angle.}

The Cabibbo rotation makes any pseudoscalar eigenstate of strong
interactions a linear combination of no longer two but, for the charged
states, of the four $({\Bbb S}^0, \vec {\Bbb P})({\Bbb D}_i), i=1\cdots 4$,
and for the neutral states, of even more since $({\Bbb P}^0, \vec {\Bbb
S})$ representations have to be included too.

Since the coefficients of the linear combinations that determine the charged
states are all real, the presence of both $\Phi_4$ and $\Phi_i, i=1,2,3$ makes
impossible the alignment of strong and electroweak eigenstates in such a way
that, inside a given representation, the charged ${\Bbb P}^\pm$ are related 
by charge conjugation. 

Let us indeed consider, for example, the two quadruplets
\be
\Xi_1 ={1\over 2i}(c_\theta(\Phi_1 + \Phi_2) -s_\theta (\Phi_3 + \Phi_4))
\ee
and
\be
\Xi_2 = {1\over 2i}(c_\theta(\Phi_1 + \Phi_2) -s_\theta (\Phi_3 - \Phi_4)).
\ee
Their charged states are
\bea
\Xi_1^+ &=& \pi^+,\cr 
\Xi_1^- & & c_\theta^2 \pi^- 
                   +c_\theta s_\theta(K^- -D^-) -s_\theta^2 D_s^-,\cr
\Xi_2^+ &=& c_\theta^2 \pi^+ +c_\theta s_\theta(K^+ -D^+) -s_\theta^2 D_s^+,\cr
\Xi_2^- &=& \pi^-;
\eea
so, even $\pi^\pm$ cannot be now electroweak mass eigenstates.

In the neutral sector, the ``strong'' $\pi^0$ can be written

\vbox{
\bea
\pi^0 &=& \left.{(\bar u u - \bar d d)\over\sqrt{2}}\right|_{P-odd}
= {1\over 2}({\Bbb P}^0 -i {\Bbb P}^3)({\Bbb D} =\left(\ba{cc} 1 & 0 \cr
                                                 0 & 0 \ea\right))\quad
       -{1\over 2}({\Bbb P}^0 +i {\Bbb P}^3)({\Bbb D}=\left( \ba{cc} 
                                         c_\theta^2 & -c_\theta s_\theta\cr
                                        -c_\theta s_\theta & s_\theta^2
                                                       \ea\right))\cr
&=& {1\over 2}\left(
     -i ({\Bbb P}^3({\Bbb D}_1) + {\Bbb P}^3({\Bbb D}_2))
     +c_\theta s_\theta ({\Bbb P}^0 +i {\Bbb P}^3)({\Bbb D}_3)
   + s_\theta^2 ({\Bbb P}^0 +i {\Bbb P}^3)({\Bbb D}_2)
\right);
\eea
}
because of the presence of the ${\Bbb P}^0$'s it is now connected, by the 
action of the ${\Bbb T}^\pm$ generators of the gauge group (\ref{eq:action}), 
not only to pseudoscalar charged particles but to charged scalars.
Consequently, aligning the strong and electroweak neutral pions
unavoidably  leads to charged electroweak mass eigenstates that are
mixtures of scalars and pseudoscalars.

Since we set aside this possibility from the beginning,
\footnote{We rejected it for the sake of simplicity, not because it is
uninteresting: it is left here as an open question, to be investigated in
further works.}
we see that,  as soon as the Cabibbo rotation is operating, there is no way
either for the $\pi^0$ to be an electroweak mass eigenstate.
 
Note that a mass term $m^2 \Xi_1\otimes \Xi_1$ is not $CP$ invariant unless
the one for $\Xi_2$ has the same coefficient;
a crossed mass term $ m^2 \Xi_1\otimes \Xi_2$ is $CP$ invariant.

The situation in the $K$ and $D$ sector is now even more intricate
than in the case of a vanishing Cabibbo angle. The same construction as in
the previous section can lead to electroweak mass eigenstates which are not
$CP$ eigenstates, but these states now involve $\pi$ and $D_s$ mesons too,
which cannot be disentangled from $K$ and $D$: in general, $CP$ violation
cannot be restricted to the sole sector of $D$ and $K$ mesons.

\section{Conclusion.}

It it essentially made of open questions.

The electroweak standard model for quarks is simple and extremely seducing
but:\l
-\quad it could be that it  only  describe a limited set among
all the features of mesons and thus be too restrictive;\l
-\quad if one adds to it a gauge theory of quarks and gluons 
\cite{MarcianoPagels}, 
the missing features could be thought to be hidden in the 
process of ``confinement''; but we are unable to solve it, and the remarks
of Feynman at the beginning of his $1981$ paper \cite{Feynman}
are still topical.

In general, going to smaller and smaller substructures  aims at a
greater simplicity; this may however  not be optimal, in particular when 
the gap deepens between the notion of ``fields'' and ``particles'', that is
between what we compute and what we observe.

We have given in previous works \cite{Machet1,Machet2,Machet3}
and here an example of a  gauge theory for $J=0$ mesons which is not only
compatible with the $SU(2)_L \times U(1)$ standard model for quarks, 
but also richer without invoking quantum chromodynamics; it cannot 
pretend, of course, to describe mesons of higher angular momenta, for which
compositeness is certainly appealing (though a Regge-like behaviour
\cite{Regge} has not yet been attached to quantum chromodynamics either). 
But famous examples  remind
us that different descriptions of the same reality should not be thought 
to exclude each other but to concur towards a better understanding of
observed phenomena.

A very pressing question clearly concerns what is detected and measured in
experiments, where all data are analyzed through the ``filter'' of a 
theoretical model; good compatibility with a given model does not exclude 
a better filter which would still improve the agreement. In this respect, I
suggested in \cite{Machet3} that the custodial $SU(2)_V$ described there
as directly related to the quantization of the electric charge 
could be found to be an exact symmetry if the data were analyzed not with
the ``filter'' of the standard model for quarks, but with an electroweak 
gauge theory
for mesons, in which the internal lines in the perturbative diagrams are 
also the propagators of asymptotic states.

The suggestion made in the present work is that the nature of observed mesonic
mass eigenstates
may not yet be so well understood, specially as far as electroweak
interactions are concerned: we have seen that, unlike what is expected, the
latter strongly alter the $SU(4)$ classification of eigenstates;
the mixture of scalars and pseudoscalars is left, too, as an open question.

Our understanding of  $CP$ violation is also modified by the point of view 
developed here: that it can take place for two generations
demonstrates how much  our description of reality depends on the model 
that we use to interpret the experimental data.

\vskip 2cm 
\begin{em}
\underline {Acknowledgments}: it is a pleasure to thank J. Avan for his
careful reading of the manuscript.
\end{em}

\newpage\null
\begin{em}

\end{em}

\end{document}